\documentclass[useAMS,usenatbib,usegraphicx]{mn2e}
\usepackage{aas_macros}
\usepackage{amssymb}
\usepackage{amsmath}

\title[The correlation of Auger UHECRs with nearby galaxies]{The correlation of UHECRs with nearby galaxies in the Local Volume}
\author[A.~J. Cuesta \& F. Prada]{Antonio J. Cuesta$^{1,2}$\thanks{E-mail: ajcv@iaa.es} and Francisco Prada$^1$\\
$^1$Instituto de Astrof{\'\i}sica de Andaluc{\'\i}a (CSIC),
Camino Bajo de Hu\'etor 50, E-18008 Granada, Spain \\
$^2$Visiting Scholar, Institute for Theory and Computation, Harvard-Smithsonian Center for Astrophysics, 
Cambridge, MA 02138
}

\begin{document}
\bibliographystyle{mn2e}
\maketitle

\begin{abstract}
We explore the possibility of a local origin for ultra high energy cosmic rays (UHECRs). Using the catalogue of Karachentsev et al. including nearby galaxies with distances less than 10Mpc (Local Volume), we search for a correlation with the sample of UHECR events released so far by the Pierre Auger collaboration. The counterpart sample selection is performed with variable distance and luminosity cuts which extract the most likely sources in the catalogue. The probability of chance correlation after penalizing for scans is 0.96\%, which corresponds to a correlation signal of $2.6\sigma$. We find that the parameters that maximize the signal are $\psi=3.0^{\circ}$, $D_{\rmn{max}}=4$Mpc and $M_{\rmn{B}}=-15$ for the maximum angular separation between cosmic rays and galaxy sources, maximum distance to the source, and sources brighter than $B$-band absolute magnitude respectively.  This implies a preference for the UHECRs arrival directions to be correlated with the nearest and most luminous galaxies in the Local Volume, while the angular distance between the cosmic ray events and their possible sources is similar to that found by The Pierre Auger Collaboration using active galactic nuclei (AGNs) within 70-100Mpc instead of local galaxies. We note that nearby galaxies with $D<10$Mpc show a similar correlation with UHECRs as compared to well-known particle accelerators such as AGNs, although less than 20\% of cosmic ray events are correlated to a source in our study. However, the observational evidence for mixed composition in the high-energy end of the cosmic ray spectrum supports the possibility of a local origin for UHECRs, as CNO nuclei can travel only few Mpc without strong attenuation by the GZK effect, whereas the observed suppression in the energy spectrum would require more distant sources in the case of pure proton composition interacting with the CMB.
\end{abstract}

\begin{keywords}
cosmic rays -- methods: statistical
\end{keywords}

\section{Introduction}

Ultra-high energy cosmic rays (UHECRs) may become a complementary probe of some astrophysical objects in addition to observations in multiple wavelengths. In fact, the detection of these particles with enough statistics would represent the awake of the development of multi-messenger astrophysics. The Pierre Auger observatory in Malarg\"ue, Argentina (see \citealt{2004NIMPA.523...50A}) has devoted a large effort in the construction of an unprecedented array of water {\v C}erenkov tanks covering an area of $\sim$3,000 km$^2$, together with four fluorescence telescopes which allow for increased accuracy in energy measurements. This effort has proven to be fruitful as the first relevant results were obtained even before the completion of the entire experiment array. In particular, the unprecedented statistics on the detection of ultra-high energy cosmic rays above 10EeV=$10^{19}eV\simeq 1J$ with arrival directions measured with an accuracy better than $1^{\circ}$ has allowed to search for possible astrophysical sources of these particles with higher reliability than previous experiments (e.g. \citealt{2004APh....21..359F}). 

By the end of 2007, the Pierre Auger Collaboration concluded that the arrival directions of UHECRs above 55EeV correlate with the positions of nearby active galactic nuclei (AGN), or other objects which trace, in the same way as AGNs, the Large Scale Structure of the Universe (\citealt{2007Sci...318..938T}, \citealt{2008APh....29..188T}). Statistics of these cosmic ray events has doubled since then (from 27 to 58 events), and yet the correlation with these objects has not strengthened as compared to previous estimations \citep{2009arXiv0906.2347T}. This supports the fact that this scenario is not the only one which can reproduce the observed data, although increasing statistics of UHECR events may be able to finally disentangle their real origin. Expectations in the cosmic-ray community are optimistic, and as stated in \citet{2009astro2010S.225O}, this issue may find a solution in the next decade.

Indeed, there has been a plethora of studies in the literature trying to find out a correlation of the UHECRs with other candidate sources such as particular objects like galaxy clusters, GRBs or other potential hosts of energetic phenomena (e.g. \citealt{2005astro.ph..7679P}, \citealt{2008PhRvD..78b3005M}, \citealt{2008MNRAS.390L..88G}) and structures like the Supergalactic Plane \citep{2008arXiv0805.1746S}. Some of these works did report a similar significance as compared to the Auger result on the correlation with the angular positions of nearby AGNs. These similar values of the probability of chance correlation failed to discriminate with enough robustness between different cosmic-ray acceleration sites. In order to get stronger conclusions from this scarce set of UHECR events, there have been several efforts on the development of new methods to estimate the chance probability of correlation in the case of isotropic flux with potential sources (\citealt{2009JCAP...07..023A}, \citealt{2009arXiv0906.2347T}). Despite all these efforts, there is no preferred candidate source yet.

Therefore, any complementary information from other measurements, like the shape of the energy spectrum, or the composition of these cosmic rays, plays now a decisive role. For example, interactions with CMB photons via the Greisen-Zatsepin-Kuzmin effect (\citealt{1966PhRvL..16..748G}, \citealt{1966JETPL...4...78Z}) are expected to decimate the statistics of the UHECRs arriving to the Earth as a function of their composition and travel distance (e.g. \citealt{2005NuPhS.138..465C}, \citealt{2009astro2010S.225O}). Recent claims of a detection of this GZK cutoff (\citealt{2008PhRvL.100j1101A}, \citealt{2008PhRvL.101f1101A}) raises the question of how far away these cosmic accelerators must be located in order to present the observed suppression. Besides, simple considerations regarding the coincidence of the GZK and UHECR clustering energies point to the fact that the composition of UHECRs must deviate from pure proton composition and tend to be CNO--like \citep{2008JPhCS.120f2006D}, which is favoured by observations (\citealt{2009arXiv0901.3389B}, \citealt{2009arXiv0902.3787U}, \citealt{2009arXiv0906.2319T}). As CNO nuclei can travel shorter distances as compared to protons, this sets a strong limit on the distance of UHECR sources, without suffering from the large deflection angles by the galactic magnetic field in the case of iron nuclei \citep{2009arXiv0909.1532T}.

In this paper we address the possibility of a local origin for the Auger UHECRs. We use the only data published so far in \citet{2008APh....29..188T}, which comprises 27 UHECR events detected prior to 1 September 2007, with energies above 57EeV (which has been revised to 55EeV due to new calibration procedures). This energy scale turns out to be however very interesting in the search for these comic ray sources, as \citet{2009arXiv0901.4699M} showed that the clustering above this energy is statistically significant. Most of the recent studies, including \citet{2008APh....29..188T}, have focused so far on sources located as far as $~75-100$Mpc, which corresponds to the GZK horizon for protons. Here we assess if there is any correlation signal with nearby objects with distance $D<10$Mpc (Local Volume) which is roughly the GZK horizon for CNO nuclei. In order to explore this hypothesis, we take advantage of the Local Volume catalogue of galaxies of \citet{2004AJ....127.2031K}, which is about 70--80\% complete up to 8Mpc. This allows us to evaluate the likelihood of galaxies in the Local Volume as UHECR accelerator sites.

This paper is organized as follows: in Section~\ref{sec:methodology} we describe the methodology used in our analysis and the catalogue of nearby galaxies in the Local Volume. In Section~\ref{sec:results} we present the results of the chance probability of the correlation of UHECRs with these objects. We discuss our results and conclude in Section~\ref{sec:conclusion}.

\section{Methodology}
\label{sec:methodology}

\begin{figure}
\begin{center}
\includegraphics[width=0.5\textwidth]{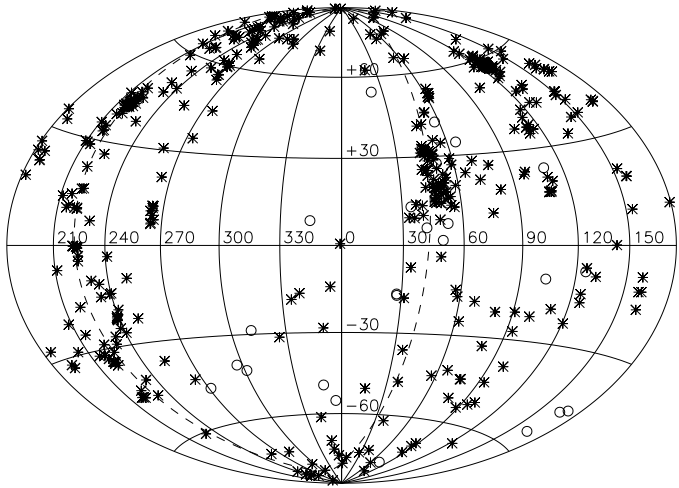}
\end{center}
\caption{The spatial distribution in galactic coordinates of cosmic-ray arrival directions (circles) and nearby galaxies from the catalogue of \citealt{2004AJ....127.2031K} (stars) shown in Aitoff projection. The Supergalactic plane is also shown with dashed line for reference.}
\label{fig:pos}
\end{figure}

In this paper we compare the distribution of UHECR arrival directions from \citet{2008APh....29..188T} with the positions on the sky of nearby galaxies from the catalogue of \citet{2004AJ....127.2031K}. This catalogue comprises 451 neighbouring galaxies up to an estimated distance of $D\lesssim 10$Mpc or a radial velocity $V_{\rmn{LG}}<550 {\rmn {km~s}}^{-1}$ with respect to the Local Group centroid. In Figure~\ref{fig:pos} we represent an Aitoff projection of the angular distribution of cosmic-ray arrival directions (circles) together with the locations of all the galaxies in the whole \citet{2004AJ....127.2031K} Local Volume catalogue (stars), both in galactic coordinates. The Supergalactic plane is also shown with dashed line for reference. As expected, a large number of nearby galaxies in the catalogue are clustered near the Supergalactic plane. This makes these particular sources good candidates for cosmic ray accelerators, as the arrival directions of UHECRs seem to be correlated with this plane (see \citealt{2008arXiv0805.1746S}). In particular, there are a large number of these UHECR events which are clustered near the Supergalactic plane at the location of Centaurus A, as previously pointed out by \citet{2008APh....29..188T} and confirmed in \citet{2009arXiv0906.2347T}, although some studies conclude that this region is unlikely to be a significant source of UHECRs \citep{2009MNRAS.395.1999C}.

The information available in the catalogue of \citet{2004AJ....127.2031K} includes the absolute magnitude in the $B$-band and real distance information of the galaxies therein. This allows us to select the nearest and most luminous galaxies, which might be associated to the highest UHECR flux provided that there is a correlation between $B$-band luminosity and cosmic ray acceleration power. Hence, the relevant quantities we choose to conform our parameter space are: the angular separation $\psi$ between cosmic-ray events and galaxy sources, the threshold in the distance to the candidate sources $D_{\rmn max}$, and the maximum absolute magnitude in $B$-band $M_{\rmn{B}}$. We did not study any dependence in energy of cosmic-ray events, as the data released by the Auger collaboration in 2007 only included 27 events, which is too sparse to make sub-samples with different energies. We show in Table~\ref{tab:scan} the values in the parameter space explored in our study for these three parameters $\psi$, $D_{max}$ and $M_{\rmn{B}}$. We note that the most finely scanned quantity is the angular separation $\psi$, which is related to the deflection angle by the intervening magnetic fields. The small number of galaxies in the catalogue of \citet{2004AJ....127.2031K} do not allow to perform a fine slicing in all the three parameters simultaneously, so we choose $\Delta\psi=0^{\circ}.1$ for comparison with Fig.~3 in \citet{2008APh....29..188T}, and took several slices in distance and luminosity. The sampled values according to Table~\ref{tab:scan} extends over a grid of only 5,325 points in this space, which is computationally inexpensive.

\begin{table}
\caption{Parameter space explored in this study. We show the minimum and maximum values, $X_{\mathrm{min}}$ and $X_{\mathrm{max}}$, as well as the scanning step $\Delta X$, for the three parameters explored in our study, i.e., $\psi$, $D_{max}$ and $M_{\rmn{B}}$(see text for details).}
\begin{center}
\begin{tabular}{c|ccc}
Quantity $X$ & $\psi$ ($^{\circ}$)      & $D_{\mathrm{max}}$ (Mpc)  &  $M_{B,\mathrm{max}}$\\
\hline
$X_{\mathrm{min}}$ & $1.0$   & $3.0$   & $-18$  \\
$X_{\mathrm{max}}$ & $8.0$   & $10.0$  & $-6$   \\
$\Delta X$& $0.1$   & $0.5$   & $3$    \\
\end{tabular}
\end{center}
\label{tab:scan}
\end{table}

In order to have an estimation of the probability of rejecting the null hypothesis (i.e. that the UHECRs in our sample show a distribution compatible with isotropy), we need to calculate the probability of correlation with a possible isotropic flux, following \citet{2008APh....29..188T}. Provided that a particular combination of values for our three parameters can show a pronounced minimum in this probability, we have to explore the full parameter space so that we can study the case in which isotropy seems to be unlikely. Therefore, for each set of values of our parameters we evaluate the probability of isotropic flux associated to (at least) $N_{\rmn{corr}}$ correlated source--cosmic ray event pairs. This is given as expected by the binomial distribution:
\begin{equation}
P_{\rmn{data}}=\sum_{j>N_{\rmn{corr}}}^{N_{\rmn{tot}}}\binom{N_{\rmn{tot}}}{j}p^j(1-p)^{N_{\rmn{tot}}-j}
\label{eqn:binom}
\end{equation}

where $p$ is the probability that an event drawn from isotropy correlates with a galaxy within $\psi$, using the sub-sample given by the other two parameters. To make this more clear, let us assume that the observatory has uniform exposure for the entire sky. In this case the circles in the sky of radius $\psi$ around galaxies in the catalogue which are brighter than $M_{\rmn{B}}$ and nearer than $D_{\mathrm{max}}$ define an area which is a fraction $p$ of the full $4\pi$ sr sky. The value of $p$ is estimated by MonteCarlo sampling of the sphere, calculating the fraction of points whose distance to a galaxy in the catalogue is smaller than a given angle $\psi$. For very small values of $\psi$, this estimation is straightforward as the areas covered by individual galaxies do not overlap, and hence the total area is equal to the number of galaxies $N$ times the area covered by the spherical cap subtended by the angle $\psi$, i.e. $0.5N(1-\cos\psi)$. On the other hand, for large values of $\psi$ or large number of galaxies, the area around individual galaxies will overlap and this approximation is no longer valid, so a numerical estimation is needed. In the general case, we also need to take into account the declination dependence on the exposure due to partial sky coverage by the observatory. This is done by generating events according to the relative exposure distribution $\omega(\delta)$. The number of points we used in the MonteCarlo sampling of the sphere is $10^6$ so we have a $10^{-3}$ uncertainty on the determination of $p$ for each node in the grid of the parameter space.

\section{Results}
\label{sec:results}

\begin{figure*}
\includegraphics[width=0.33\textwidth]{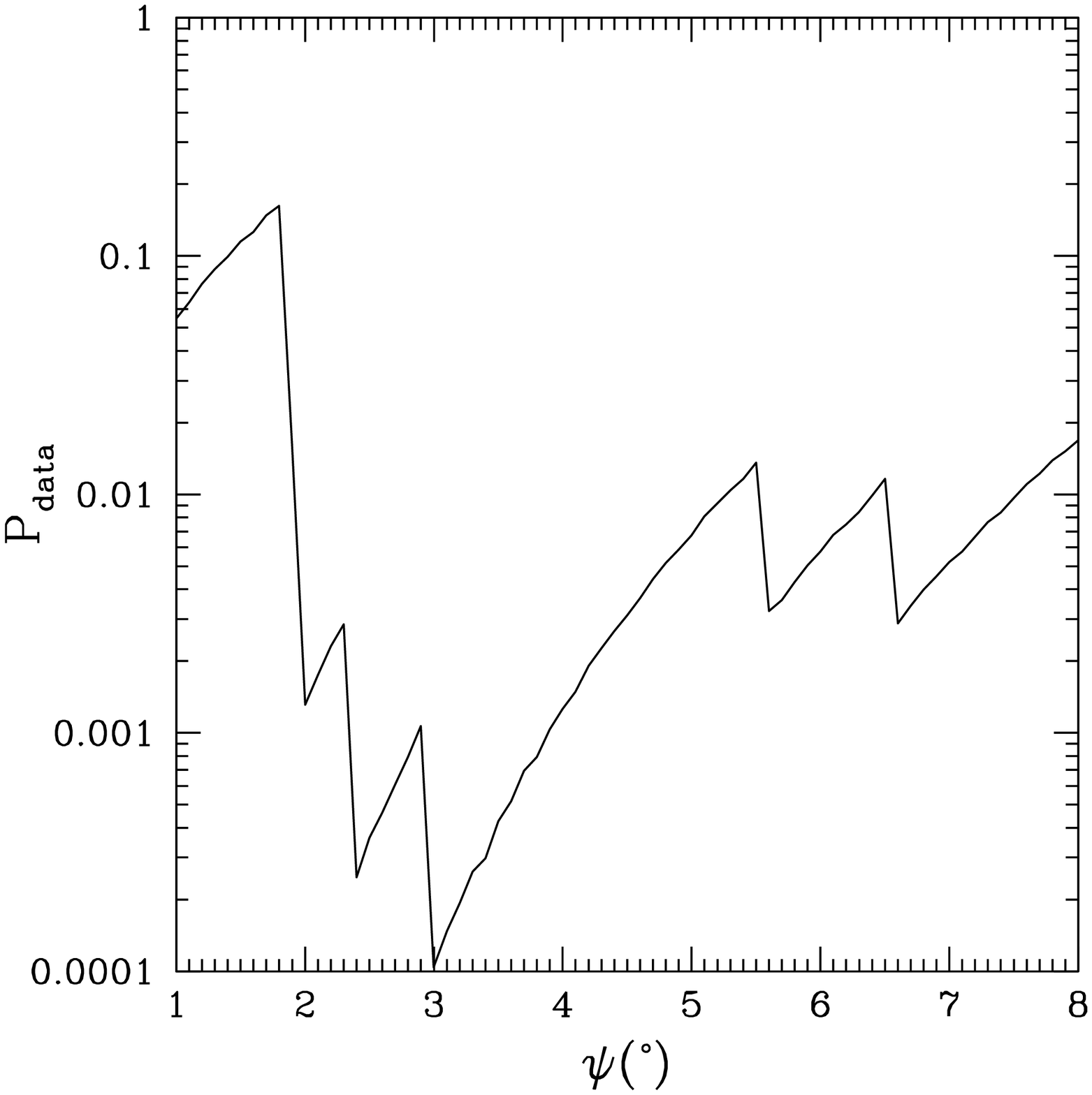}
\includegraphics[width=0.33\textwidth]{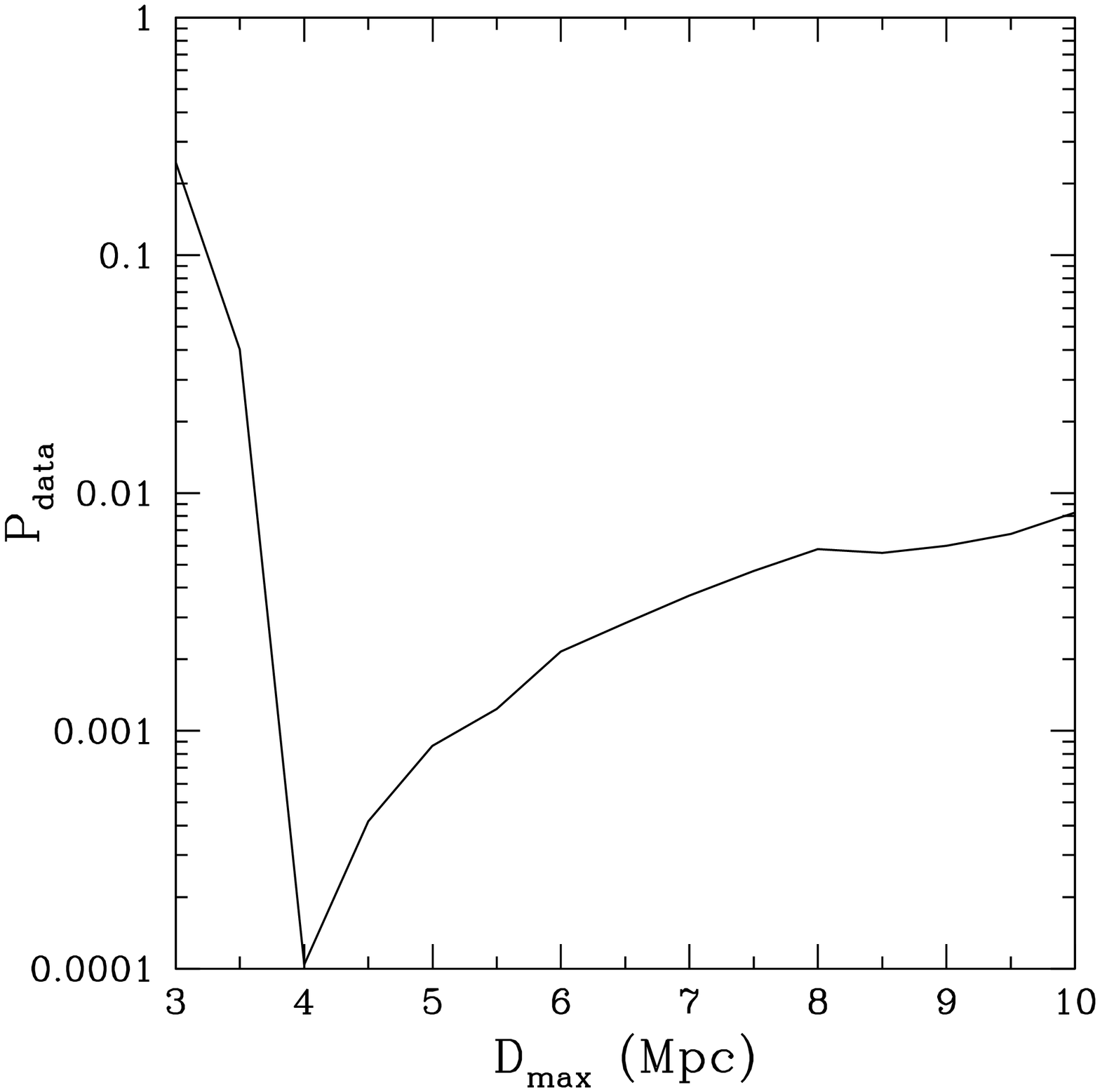}
\includegraphics[width=0.33\textwidth]{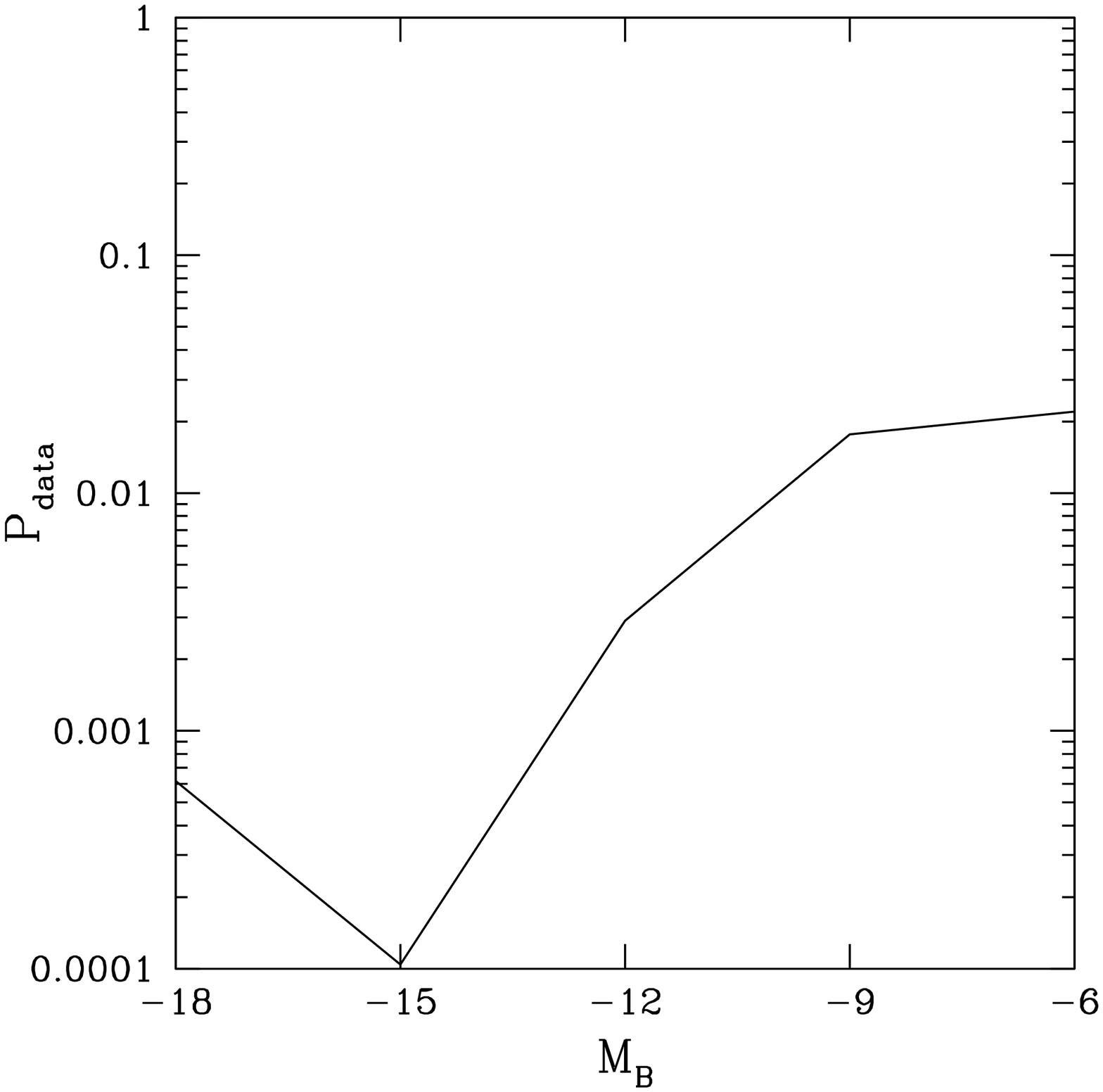}
\caption{The chance correlation probability as a function of each one of the scan parameters, holding the other two fixed at their minimum of $P_{\rmn{data}}$. \textit{Left panel}: Angular separation $\psi$. \textit{Middle panel}: Maximum galaxy distance $D_{max}$. \textit{Right panel}: Maximum absolute magnitude in the $B$-band.}
\label{fig:scan}
\end{figure*}

We now apply the method described in the previous Section to calculate the probability in the case of isotropic flux of $N_{\mathrm{corr}}$ correlated source--cosmic ray event pairs, as a function of the parameters $\psi$, $D_{max}$ and $M_{B}$. A full exploratory scan is performed over the entire parameter space, and as a result we find that this probability reaches its absolute minimum (hereafter $P_{\rmn{data}}$) at cosmic ray event deflections below $\psi=3.0^{\circ}$ and galaxies accelerating UHECRs with distances below $D_{\rmn{max}}=4$Mpc and brighter than $M_{\rmn{B}}$=-15. This implies a preference for the UHECRs arrival directions to be correlated with the nearest and most luminous galaxies in the catalogue, while the angular distance between the cosmic ray events and their possible sources is similar to that found by \citet{2008APh....29..188T} using AGNs instead of local galaxies. In order to visualize the dependence of this probability with individual parameters, we refer to Figure~\ref{fig:scan}. Here we display the chance probability of correlation with the Auger UHECRs above $E>55$EeV as a function of each one of the scan parameters, keeping the other two parameters fixed at the absolute minimum of the probability of isotropic flux. When the angular distance $\psi$ varies, we find several local minima, although the region in which the chance correlation probability is below 0.1\% is in the range of 2--4$^{\circ}$. On the other hand, the behaviour with the other two scan parameters is quite robust with a single minimum, although this is also due to the coarseness in the scanning values in this case. Nevertheless, it seems clear that including distant and intrinsically faint sources decreases the correlation signal. Moreover, given the value of $\psi$ the deflection of cosmic ray trajectories due to intervening magnetic fields is not large, suggesting that these events are light nuclei, as suggested by \citet{2008APh....29..188T}.

As shown in Figure~\ref{fig:scan}, the probability at the absolute minimum is $P_{\rmn{data}}=1.1\times 10^{-4}$. This signal is similar to the previous result of \citet{2008APh....29..188T}, who found $P_{\rmn {data}}=2\times10^{-4}$, and even stronger than the revised result by \citet{2009arXiv0906.2347T}, who found that $P_{\rmn {data}}=6\times10^{-3}$ for the correlation with AGNs up to $75$Mpc. Nevertheless, our result is not free from the negative impact of trial factors in a posteriori anisotropy searches, as it was not tested on an independent data set with the scan parameters specified a priori. For reference, the Auger result including all the 27 events is $P_{\rmn {data}}=4.6\times10^{-9}$. Also, our value is very sensitive to the low number of correlating events at this minimum (five cosmic rays associated to four sources), so removing one UHECR can boost $P_{\rmn{data}}$ to $1.10\times 10^{-3}$. However, it is important to note that Eq.~(\ref{eqn:binom}) is not the true probability of chance correlation due to isotropic flux, as pointed out by \citet{2004APh....21..359F}. Although this exploratory scan is useful in order to detect the location of the strongest potential correlation signal, the fact that we have chosen our parameters in such a way to get the minimum value of this probability has to be taken into account. In order to correct the value $P_{\rmn{data}}$ that we have just obtained, we need to include a probability penalty due to parameter selection. This is fixed by performing $n_{\rmn{MC}}$ MonteCarlo realizations of the cosmic-ray data set, which are drawn from a isotropic distribution but they are also selected in such a way that they mimic the events which the Pierre Auger Observatory might detect, due to declination dependence of the exposure. Therefore, the mock cosmic ray data set assume isotropic flux but the probability of a given arrival direction ($\alpha,\delta$) is proportional to the exposure function of the Auger observatory on that position (see e.g.~\citealt{2006JCAP...01..009C}).

The isotropic samples are obtained as usual by generating random points on the sphere, i.e. picking uniformly distributed random numbers in $\sin\delta$ and $\alpha$, so that the number density of points is completely homogeneous over the solid angle unit $\mathrm{d}\Omega = \cos\delta\mathrm{d}\delta\mathrm{d}\alpha$. To account for the Auger observed sky, we accept a random point on the sphere if an uniformly distributed random number between 0 and 1 is smaller than the relative exposure $\omega$ normalized by its maximum value $\omega_{\rmn{max}}$, otherwise this point is rejected. According to \citet{2001APh....14..271S}, the relative exposure is given by:
\begin{equation}
\omega(\delta) \propto \cos(\lambda)\sin(\alpha_m(\zeta))\cos(\delta)      +\alpha_m(\zeta)\sin(\lambda)\sin(\delta)
\end{equation}
where $\lambda \simeq -35^{\circ}$ is the latitude corresponding to the location of the Pierre Auger Observatory, $\zeta$ is given by:
\begin{equation}
\zeta\equiv \frac{\cos(\theta_{\rmn{max}})-\sin(\lambda)\sin(\delta)}{\cos(\lambda)\cos(\delta)}
\end{equation}
($\theta_{\rmn{max}}=\pi/3$ is the cut in zenith angle $\theta$ of the arrival directions applied by Auger), and $\alpha_m$ is defined as follows:
\begin{equation}
\alpha_m=
\left\{
\begin{array}{ll}
\pi & \rmn{if}~\zeta < -1 \\
\arccos \zeta & \rmn{if}~ -1 < \zeta < +1 \\
0 & \rmn{if}~\zeta > +1
\end{array}
\right.
\end{equation}

Therefore, if we perform the same exploratory scan in the same way as in the beginning of this Section, for each different set $i$ of Monte Carlo generated arrival directions $(i=1,\ldots,n_{\rmn {MC}})$, we can just count the number of trials $n^*_{\rmn{MC}}$ for which the absolute minimum (over the entire parameter space) of the probability of isotropic flux is lower than the actually observed minimum, $P^i_{\rmn{min}} \leq P_{\rmn{data}}$. The chance probability of observing $P_{\rmn{data}}$ is \citep{2004APh....21..359F}:
\begin{equation}
P_{}=\frac{n^*_{\rmn{MC}}}{n_{\rmn{MC}}}
\end{equation}

We can consider this value as an estimate of the chance probability of correlation due to isotropic flux, as in \citet{2008APh....29..188T}. We performed $10^6$ Monte Carlo realizations of the Auger cosmic ray data set, in order to keep our uncertainties below 0.1\%. The value we obtain for this probability after correcting for parameter selection is the following one:
\begin{equation}
P_{\rmn {chance}}=(9.6\pm 1.0)\times 10^{-3}
\end{equation}

We find that the probability of chance correlation of the highest energy ($E>55$EeV) Auger events with nearby galaxies from the \citet{2004AJ....127.2031K} catalogue is around 0.96\%. This small probability corresponds to a $2.6\sigma$ correlation between both samples. We point out that more than the half of the 27 Auger events above $E>55$EeV are within $4^{\circ}$ of a nearby ($D<10$Mpc) galaxy, which made it worth of exploring. This result goes along the same direction as the previous studies, finding a hint of violation of isotropy in the Auger sample.

There are 5 out of 27 correlating events, while only 0.5 should be expected on average in the case of isotropic flux, as the effective fraction of the sky for the values of the parameters at the minimum is $p=0.018$. Of course the number of correlating events is much lower than the value found by \citet{2008APh....29..188T} (20 out of 27, with 5.6 expected in the case of isotropic flux), but it should be considered that the effective fraction of the sky covered (at the minimum $P_{\rmn{data}}$) by 442 objects in the V{\'e}ron-Cetty AGN catalogue with $z\leq 0.017$ is much larger ($p\simeq 0.21$) than the 31 objects with our parameters ($p=0.018$).

The galaxies which correlate with the arrival directions of the cosmic rays, are NGC300 (spiral galaxy type SA), NGC4945 (Sy2), NGC5102 (HII) and NGC5128 (Sy2). With exception of NGC300, they mostly show nuclear activity and both Seyfert galaxies are included in the AGN catalogue used by \citet{2008APh....29..188T}. In particular NGC5128, which is associated with 2 cosmic-ray events, is inside the well known Cen~A region. Thus, it might happen that the correlation we find could just be inherited by the correlation found by the Auger Collaboration if the latter is finally confirmed. This is reasonable, as only a few tens of AGNs from the V{\'e}ron-Cetty catalogue are located at a distance of $D<10$Mpc, but it is important to keep in mind that CNO UHECRs from larger distances can be heavily suppressed.

\section{Discussion and conclusion}
\label{sec:conclusion}
We have explored the possibility of the correlation of nearby galaxies from the catalogue of \citet{2004AJ....127.2031K} with ultra high energy cosmic ray events from the Pierre Auger Collaboration. This analysis is highly motivated in the light that a mixed CNO (as opposed to pure proton) composition of UHECRs, which is the preferred scenario indicated by current data, could not survive the GZK cutoff unless the sources are located in the Local Volume. Therefore, any correlation with these nearby sources would explain both composition and correlation at the same time. The analysis of the data concludes a $2.6\sigma$ correlation between both samples. There are 5 out of 27 correlating events, while 0.5 should be expected on average in the case of isotropic flux. Due to the small number of correlating events, we cannot specify the scan parameters a priori, thus making our results sensitive to the negative impact of a posteriori anisotropy searches. The probability of chance correlation due to isotropic flux after penalizing for scans is $P_{\rmn{chance}}=0.96\times 10^{-2}$, rejecting the hypothesis of isotropy of arrival directions of UHECRs at 99\% confidence level. For comparison, the same result before penalizing for scans is $\sim 1\times 10^{-4}$ which is similar to the $\sim 2\times 10^{-4}$ result (with scan parameters set a priori) from the correlation found by \citet{2008APh....29..188T} with the AGNs in the V{\'e}ron-Cetty catalogue, and stronger than the revised value of $\sim 6\times 10^{-3}$ when they include the new UHECR events from 1 September 2007 to 31 March 2009 \citep{2009arXiv0906.2347T}. This is an indication that there is no clear source of the UHECRs, and finding a lower value of the probability of chance correlation does not guarantee that this is actually the preferred source as compared to other sources which present similar values of $P_{\rmn{chance}}$. This was already noticed even in early papers such as \citet{2001JETPL..74..445T}, and in any case it seems unlikely to get a $>3.3\sigma$ detection \citep{2008JCAP...05..006K}. Interestingly, most of the objects in the catalogue of \citet{2004AJ....127.2031K} which correlate with Auger events, are already included in the AGN catalogue, which has around 30 objects with $D<10$Mpc. This may indicate that the relation between nearby galaxies and UHECRs found in our work could be inherited from the \citet{2008APh....29..188T} result. In any case, the so-called AGN hypothesis has proven to be controversial (\citealt{2008JETPL..87..461G}, \citealt{2009ApJ...693.1261M}), stressed by the deficit of events from the direction of the Virgo supercluster. On the other hand, the correlation with the distribution of luminous matter seems to be well established \citep{2008JCAP...05..006K}, without making any reference to any particular source which can accelerate particles up to these energies.

The small number of correlating events with nearby galaxies might also point to the possibility of the absence of UHECR accelerators in our local environment, although there is no definitive answer yet. Surprisingly, most of the objects in the \citet{2004AJ....127.2031K} catalogue which correlate with the UHECR sample (with the exception of NGC300) are located in a small region in the sky, with Supergalactic coordinates $150^{\circ}<SGL<165^{\circ}$ and $-10^{\circ}<SGB<-5^{\circ}$. This region is near the Supergalactic plane, pointing to preferential arrival directions from nearby galaxies located in a high density region. Future measurements and hopefully the next data release with a total of 58 events as announced in \citet{2009arXiv0906.2347T} will help to clarify the origin of nearby and distant UHECRs, which will allow to study more in detail the acceleration physics behind them.\\

The authors acknowledge support from the Spanish MEC under grant PNAYA 2005-07789. A.J.C. thanks support of the MEC through Spanish grant FPU AP2005-1826. We especially want to thank Vasiliki Pavlidou from CalTech Astronomy Department for useful comments on this paper. This work made extensive use of the x4600 machine at CICA (Spain). We thank Claudio Arjona and Ana Silva at CICA for making this machine available to us.

\bibliography{mycites}

%\bsp
\end{document}